%% file: main.tex
\documentclass[twocolumn]{aastex631}

\AtBeginDocument{}
\usepackage{amsmath}
\usepackage{graphicx}
\graphicspath{{./}{figures/}}
\usepackage{xcolor}
\usepackage[normalem]{ulem}
\usepackage{hyperref}%

\input{figures.tex}
\begin{document}
\title{Feedback-Conditional 3D Reconstruction of Cosmic Baryons with Flow Matching}

\author[0000-0001-9580-1043]{Maryam Hussaini}
\email{maryam.hussaini@cfa.harvard.edu} %
\affiliation{Center for Astrophysics $\mid$ Harvard \& Smithsonian, Cambridge, MA 02138, USA}

\author[0009-0004-3662-2263]{Aizhan Akhmetzhanova}
\affiliation{Department of Physics, Harvard University, Cambridge, MA 02138, USA}

\author[0000-0002-6069-2999]{Carolina Cuesta-Lazaro}
\affiliation{Center for Astrophysics $\mid$ Harvard \& Smithsonian, Cambridge, MA 02138, USA}

\author[0000-0002-7587-6352]{Liam Connor}
\affiliation{Center for Astrophysics $\mid$ Harvard \& Smithsonian, Cambridge, MA 02138, USA}

\begin{abstract}
The complex relationship between galaxies and diffuse gas at large scales depends strongly on astrophysical feedback and cosmology. The dispersion measure (DM) of Fast Radio Bursts (FRBs) is a column of cosmic baryon density and is dominated by the intergalactic medium, which is not necessarily captured by traditional halo model prescriptions. We introduce a generative model to reconstruct 3D gas density fields from galaxy surveys, including redshift space distortions. The probabilistic model is trained on the CAMELS50 simulation suite (IllustrisTNG) and learns to predict the posterior distribution of gas fields conditioned on stellar mass-weighted galaxy density, cosmological parameters, the minimum stellar mass threshold for the galaxy sample, and the baryon suppression power spectrum as a measure of baryonic feedback. Our method uses the 3D gas reconstruction to estimate per-sightline FRB DMs and their posteriors. We validate the model on test boxes by showing agreement between predicted and true DM across individual lines of sight and the overall DM distribution.
The model generalizes to a box size $\sim$\,4$^3$ times larger in volume than the training boxes and can cover the volumes spanned by cosmological FRB sightlines. The model's 3D gas density predictions vary with input power spectrum and feedback scenarios, offering an inference path to constrain feedback models from FRBs and other probes such as X-rays, the SZ effect, and weak lensing. 
\end{abstract}

\section{Introduction}
\label{sec:intro}

The distribution of cosmic baryons has remained a pressing problem in extragalactic astronomy for decades. This is due to a lack of 
direct measurements and the importance of baryons in our understanding of galaxy formation and precision cosmology. The majority 
of the baryons reside in a diffuse ionized state in the intergalactic medium (IGM) and in dark matter halos \citep{shull2012,connor2024,hussaini2025}. 
The precise distribution of the gas is modulated 
by complex astrophysical processes in galaxy evolution. In recent years, new probes of the cosmic baryons have begun 
to shed light on gas at the scales of halos and above \citep{bigwood2024, popesso2026, hadzhiyska2024, connor2024, sharma2026feedback}. Several measurements
suggest that strong AGN feedback depletes the
baryon budgets of group-scale halos \citep{popesso2026, mccarty2026}.

Supernovae, AGN, and cosmic rays inject energy into
halo gas, heating it, creating winds, and pushing the gas out of galaxies and halos, or preventing its accretion and cooling. These processes regulate star formation and the growth of supermassive black holes, leading to non-linear galaxy feedback that remains poorly constrained.

In simulations, the choice of inaccurate feedback prescriptions results in discrepancies with
observations in the stellar mass function, halo gas fractions, baryon density profiles, and temperature distributions. %
In galaxy formation, feedback regulates how baryons turn into stars, shapes the circumgalactic medium (CGM), and determines galaxy populations. %
In cosmology, feedback is the contaminant that breaks the coupling between dark matter and baryons, redistributing baryons and changing the baryonic power spectrum from that of the dark matter via suppression of power at small scales \citep{vanDaalen2011}. Understanding feedback offers a path to answering questions in galaxy formation \citep{Dekel, croton, Somerville_2015, Tumlinson17}, and alleviates a key systematic in weak lensing cosmology surveys \citep{euclid, Tr_ster_2022, amon, bigwood2024}.

\figOverview
Given that feedback and cosmology are responsible for 
the distribution of cosmic baryons, methods that allow us to infer the 3D distribution of gas are valuable for 
myriad areas in extragalactic astronomy. Reconstruction of the dark matter and velocity field from galaxy surveys has made significant advances in the past decade \citep{hong2021, park2023, Zhao2025, shi2025_velocity, wang2023_velocity}. But reconstructing the 3D gas field directly from galaxies
is less developed. In \citet{Kvasiuk2024}, they map galaxy point clouds onto both dark matter and electron overdensity with a hybrid graph and convolutional network trained on $(25\,h^{-1}\,\mathrm{Mpc})^3$ CAMELS boxes, and \citet{Krywonos2026} show that this approach is superior to linear transfer function and halo model reconstructions for both fields. Both return a single best fit field rather than a distribution, and where feedback is included it enters as astrophysical parameters in CAMELS rather than as a scale resolved quantity.
\citet{liu2025} paint optical depth and Compton-$y$ onto $N$-body simulations using a Gaussian process transfer function and hybrid effective field theory, starting from the dark matter field instead of from galaxies. No method yet returns a probabilistic 3D gas field that captures the uncertainty in the galaxy to gas mapping, explicitly conditions on the scale dependent strength of feedback, and connects to a direct observable of the diffuse gas.

A direct observable of this diffuse gas is provided by fast radio bursts (FRBs). FRBs are bright, millisecond-duration radio transients originating from galaxies across the Universe \citep{lorimer2007, petroffreview}. As the burst propagates toward us, its radio signal interacts with the ionized gas it encounters along the way and higher frequencies arrive earlier than lower frequencies, an effect known as cold plasma dispersion. The magnitude of this delay, the Dispersion Measure (DM), is directly proportional to the integrated free electron column density along the line of sight. Because it depends only on the electron column, the DM is an incisive probe of ionized gas at large scales, unweighted by metallicity, temperature, or bulk velocity. The cosmic dispersion measure to a source at redshift $z_s$ is
\begin{equation}
    \mathrm{DM}_{\rm cos}(z_s) 
    =
    \int_0^{z_s}
    \frac{c\,n_e(z)}
         {(1+z)^2 H(z)}
    \, dz ,
    \label{eq:dm}
\end{equation}
where $n_e(z)$ is the proper free electron density along the line of sight, $H(z)$ is the Hubble parameter, and $c$ is the speed of light. Averaging over many sightlines yields the mean cosmic dispersion measure:
$n_e(z)$ in Equation~\ref{eq:dm} can be replaced by the cosmic mean free electron density, $\bar{n}_e(z)$, to obtain $\left < \mathrm{DM_{cosmic}} \right >$. Because $\bar{n}_e$ is set by the cosmic baryon density, the mean DM--redshift relation directly traces the integrated ionized baryon content of the Universe \citep{macquart2020, connor2024}. An FRB signal accumulates DM contributions from the gas in its host galaxy, the intergalactic medium, the circumgalactic medium of intervening halos, and the Milky Way. However, the DM alone tells us only the total amount of gas along a sightline and not the 3D distribution of that gas.

The mapping between galaxies and the 3D gas distribution is inherently uncertain. A given galaxy field is consistent with many gas configurations, so the reconstruction should be a posterior distribution over compatible gas fields rather than a point estimate. One route to such field-level posteriors is explicit Bayesian inference, exemplified by \textsc{borg} \citep{jasche2013}. In this method, one writes down a likelihood connecting a forward modeled density field to the observed galaxies and samples the high-dimensional posterior over the initial conditions (or the evolved field) with gradient-based methods such as Hamiltonian Monte Carlo. This approach has already been applied to FRB foregrounds in the FLIMFLAM survey \citep{lee2022}. \citet{Khrykin_2024} used the \textsc{argo} algorithm \citep{ata2015, ata2021} to reconstruct the matter density field traced by real spectroscopic foreground galaxies along FRB sightlines, and combined it with the measured DMs to constrain the partition of baryons between the IGM and the CGM of intervening halos \citep{simha2020, simha2023}.

Explicit likelihood inference, however, imposes constraints on the forward model, because it must be evaluated (and, for gradient-based sampling, differentiated) at every step of the chain. The likelihood connecting galaxies to the underlying fields must be assumed explicitly, and any misspecification of that form propagates directly into the reconstruction. We take an implicit-likelihood approach, in which a generative model is trained on simulations to sample the posterior directly. Diffusion-based models have been shown to reconstruct 3D cosmic fields from observations with calibrated uncertainties \citep{mudur2022, park2023, legin2023, ono2024}, and \citet{cuestalazaro2024} demonstrated that stochastic interpolants jointly reconstruct initial conditions and cosmological parameters, reproducing the HMC posterior while requiring neither a differentiable forward model nor an explicit likelihood. 

Most relevant to this work, \textsc{BaryonBridge} \citep{baryonbridge} uses stochastic interpolants to map inexpensive particle-mesh simulations to 3D baryonic fields, conditioned on cosmology and astrophysics across the CAMELS suite. Because inference is amortized and the likelihood is learned rather than assumed, the training data can be full hydrodynamical simulations, with realistic feedback and resolution far beyond what an explicit-likelihood forward model could afford, which would be impossible to sample explicitly. We adapt the \textsc{BaryonBridge} framework to solve the inverse problem: rather than mapping dark matter to baryons, we map the observed galaxy field to the 3D gas field. We train on the second generation of the CAMELS simulation suite \citep{camels2021, genel2026}, which spans a wide range of cosmological and feedback scenarios and therefore lets the model marginalize over feedback uncertainty. The reconstruction returns a full posterior over the 3D gas distribution, making it scalable to the large samples of localized FRBs now being assembled \citep{chime2025, connor2024, shannon2024, dsa-2000-whitepaper}.

In this work, we present a generative model that reconstructs the 3D gas density field from the observed galaxy field, conditioned on cosmology and the baryon suppression spectrum, and returns a full posterior over 3D gas fields. We validate the reconstructions on the second-generation CAMELS simulations, show that the trained model generalizes to volumes far larger than those used for training ($\sim 200\,h^{-1}\,\mathrm{Mpc}$ boxes), and demonstrate how the reconstructed fields yield per-sightline FRB dispersion measures with quantified uncertainties. 

This paper is organized as follows: Section~\ref{sec:methods} describes the simulation data, preprocessing pipeline, and model architecture; Section~\ref{sec:results} presents validation results on our CAMELS test boxes and generalization to TNG300-1; Section~\ref{sec:discussion} discusses implications for FRB science and limitations; and Section~\ref{sec:conclusion} summarizes our conclusions and future directions.

\figInput 

\section{Methods}
\label{sec:methods}

We reconstruct gas density fields from redshift space galaxy density fields (Figure~\ref{fig:overview-fig}), conditioned on cosmological parameters and the baryon suppression spectrum, using the flow matching approach \citep{lipman2023flowmatching}. We train the models on the second generation CAMELS simulation suite (IllustrisTNG) \citep{genel2026}, from which we use 1000 simulation boxes each with a volume of $(50\,h^{-1}\,\mathrm{Mpc})^3$ discretized into $128^3$ voxels, a set we refer to as CAMELS50. Each box contains the 3D stellar-mass-weighted galaxy density field, the 3D gas density field, and 35 cosmological and astrophysical parameters. For a list of CAMELS50 parameters, see Table 1 in \citet{genel2026}. 

We preprocess the galaxy field by keeping galaxies above a minimum stellar mass threshold $M_{\star,\mathrm{min}}$, populating a $128^3$ voxel grid using Cloud-In-Cell (CIC) interpolation weighted by stellar mass $\log_{10}(M_{\star})$. We take $\log_{10}$ of the resulting density fields $f_{\mathrm{gal}}$ and normalize them with CAMELS50 training statistics: 
$$\delta_g = \frac{\log_{10}(f_{\mathrm{gal}}) - \mu_{\mathrm{gal}}}{\sigma_{\mathrm{gal}}},$$ 
where $\mu_{\mathrm{gal}} = -5.94$ and $\sigma_{\mathrm{gal}} = 0.586$. We vary the minimum stellar mass threshold $M_{\star,\mathrm{min}}$ during training: for each box $M_{\star,\mathrm{min}}$ is drawn log-uniformly between $10^9$ and $10^{9.5}\,M_\odot$, and $\log_{10}M_{\star,\mathrm{min}}$ is passed to the model as an additional conditioning parameter. At inference we set this threshold to $M_{\star,\mathrm{min}} = 10^9\,M_\odot$ for all results presented in this paper except Figure \ref{fig:minms}. We show an example 3D reconstruction and its predicted DMs in Figure~\ref{fig:input_output}.

The gas density field is preprocessed in a similar manner by first applying a $\log_{10}$ transformation and then standardizing to zero mean and unit variance using statistics computed from the CAMELS 50 training data ($\mu_{\mathrm{gas}} = 8.28$, $\sigma_{\mathrm{gas}} = 0.586$ for the $128^3$ resolution).

For each training box, we adopt the baryon suppression spectrum $sP(k) = P_\mathrm{hydro}(k)/P_\mathrm{DM}(k)$, the ratio of the matter power spectrum in the full hydrodynamical simulation to that of its gravity-only ($N$-body) counterpart, computed with \texttt{Pylians} \citep{villaescusa-navarro2018}. This ratio quantifies the scale-dependent redistribution of matter by baryonic physics and feedback, which suppresses power on scales $k \sim 1 - 50\,h\,\mathrm{Mpc}^{-1}$, and enhances it on smaller, galactic scales \citep{genel2026}. $sP(k)$ encodes the combined signature of cosmological parameters (which set the baryon abundance) and feedback strength (which sets the redistribution), providing a conditioning signal more sensitive than individual feedback parameters alone. We deliberately condition on $sP(k)$ rather than on the subgrid feedback parameters of the simulation suite. Parameters such as supernova and AGN efficiencies are specific to a given galaxy-formation model and do not translate between simulations, whereas $sP(k)$ summarizes how feedback has redistributed matter \citep{vanDaalen2011, sharma2024_linear_transfer, sharma2025} and is interpretable across feedback implementations, rather than being tied to the CAMELS parametrization.

The model learns conditional posterior of gas density fields $p_\theta(n_e(\mathbf{x})| \delta_g(\mathbf{x}), \, \boldsymbol{\Omega}, \, sP(k), \,\log_{10}M_{\star,\mathrm{min}})$. Here, $\delta_g(\mathbf{x})$ is the stellar-mass-weighted galaxy density field, evaluated on the same voxel grid as the gas field, with redshift-space distortions applied along the line-of-sight axis and normalized for training. The redshift-space distortions are applied to the galaxy catalog by displacing each galaxy along the line-of-sight axis by its peculiar velocity in the plane-parallel approximation before the field is deposited onto the grid. The model therefore maps a redshift space galaxy field onto a real space gas field, which is the configuration relevant to observations, where the galaxy positions are measured in redshift space but the gas column being reconstructed is in real space. The cosmological parameters the model is conditioned on are denoted by $\boldsymbol{\Omega} = \{\Omega_m,\ \sigma_8,\ H_0,\ n_s,\ \Omega_b\}$ and are varied across the CAMELS suite. $\Omega_m$ is the total matter density, $\sigma_8$ is amplitude of matter fluctuations, $H_0$ is the Hubble constant, $n_s$ is the scalar spectral index, and $\Omega_b$ is and baryon density. We choose to condition the model on this handful of parameters because they most strongly influence the large-scale matter and gas distributions. The baryon suppression spectrum sample $sP(k)$ is passed to the model as a set of 20 values at approximately logarithmically spaced $k$ modes, running from $k \simeq 0.18$ to $6.61 \, h \, \mathrm{Mpc}^{-1}$.

Flow matching learns to map a base Gaussian distribution $x_0 \sim \mathcal{N}(0, \mathbf{I})$ to the target data distribution $x_1$ by learning the velocity field $u_{\theta}(x_t)$ that transports samples from $x_0$ to $x_1$ along a linear interpolation path $x_t = (1-t)x_0 + t x_1$, where $t \in [0,1]$. The model is trained by minimizing the mean-squared error between the predicted $u_{\theta}(x_t)$ and true velocity fields $u_t = x_1 - x_0$. At inference time, the learned velocity field is integrated using a neural ODE solver with 150 steps of the second-order midpoint method. A finite step count leaves a discretization error that systematically adds power. At 150 steps our solution converges to about $2\%$. For each test box we draw 100 posterior samples.

The backbone is a 3D conditional U-Net model \citep{ronneberger2015unet} operating on $128^3$ voxel grids, with three encoder/decoder levels at channel dimensions $[32, 64, 64]$, $3\times3\times3$ circular-padded convolutions, group normalization, and SiLU activations. The galaxy field is concatenated channel-wise with the noisy gas field. The interpolant time is encoded with sinusoidal positional features followed by a two-layer MLP, while the cosmological parameters and $sP(k)$ are passed through a separate two-layer MLP; the two 128-dimensional embeddings are summed and injected into every residual block.

The model is trained with AdamW at a learning rate of $10^{-4}$, effective batch size 16 (achieved via gradient accumulation steps), $\ell_2$ gradient clipping at 1.0, and 16-bit mixed precision for 20,000 gradient steps. We use 980 boxes for training, 10 for validation, and 10 for testing. Drawing 100 posterior samples for each of the 10 test boxes takes $11.7\,\mathrm{h}$ on a single NVIDIA A100 (80\,GB), or $\simeq 42\,\mathrm{s}$ per $128^3$ posterior sample.

For our reconstructed fields, to obtain the DM, we convert the gas density to a free electron number density $n_e$ and integrate along one spatial axis of the box. We integrate along the same axis to which the redshift-space distortions were applied, so that the projection direction is the line of sight the model was trained with. Because the diffuse gas that dominates the cosmological DM is almost fully ionized, $n_e$ is proportional to the gas density, so the conversion is a constant rescaling, with the absolute normalization anchored through $\Omega_b$ (see Section~\ref{sec:discussion}).
This collapses the 3D field into a 2D map (Figure~\ref{fig:input_output}), with each pixel representing the total electron column, in $\mathrm{pc\,cm^{-3}}$, in the line of sight direction.

Our model learns a posterior over gas fields conditioned on the observed galaxy field, cosmology, the suppression spectrum, and the minimum stellar mass threshold:

\begin{equation}
n_e(\mathbf{x}_\perp, l) \sim p_\theta\left(n_e \mid \delta_g, \, \boldsymbol{\Omega}, \, sP(k), \, \log_{10} M_{\star,\mathrm{min}}\right),
\end{equation}

\noindent where $\mathbf{x}_\perp$ denotes the two sky coordinates and $l$ is the line-of-sight coordinate. The DM at each sky position is the line-of-sight integral of the electron density (Eq.~\ref{eq:dm}). Because this integral is a linear functional of $n_e$, applying it to each posterior sample of the gas field yields a corresponding sample of DM, so the gas-field posterior induces a posterior over dispersion measure directly, with no additional modeling needed.

To test generalization, we apply the trained model to TNG300-1 \citep{illustrisTNG}, the highest-resolution realization of the TNG300 volume, a simulation $\sim$4 times larger than the training boxes on a side, with a total volume $(205\,h^{-1}\,\mathrm{Mpc})^3$ vs CAMELS50's $(50\,h^{-1}\,\mathrm{Mpc})^3$. We preprocess the TNG300-1 galaxy field using the same pipeline as above, regridding the volume onto $524^3$ voxels so that the physical voxel size matches the $50\,h^{-1}\,\mathrm{Mpc}/128$ resolution of the CAMELS50 training data, and apply the same normalization statistics as during training. Because the network is fully convolutional, the entire $524^3$ volume is integrated in a single pass, with no tiling or stitching of subvolumes. At this size the activations no longer fit in GPU memory, so the reconstruction is run on CPU instead, requiring $22.5\,\mathrm{h}$ and a peak resident memory of $390\,\mathrm{GB}$ across 64 cores for a single posterior sample. All simulations are evaluated at redshift $z = 0$.

\figPatchHist

\section{Results}
\label{sec:results}
We validate the model on 10 CAMELS50 test boxes and on a larger TNG300-1 box. We show that the reconstruction is accurate, that it can be controlled through the conditioning suppression spectrum $sP(k)$, and that it extends to the large volumes needed to assemble realistic FRB sightlines.
Figure~\ref{fig:input_output} shows the model's reconstruction quality on a representative CAMELS50 test box. Although the model reconstructs 3D gas density, the FRB observable is the dispersion measure, so we sum the reconstructed gas along the line of sight to produce 2D dispersion measure maps. The reconstructed DM map recovers the large-scale structure of filaments, nodes, and voids present in the true DM map, even with sparse galaxy inputs. To quantify this, the bottom left panel of Figure~\ref{fig:input_output} compares the reconstructed and true DM sightline by sightline, as a 2D histogram over all lines of sight in the box. The scatter is clustered around the 1:1 line, with a residual scatter of $\sigma = 0.193$ dex and a best-fit slope of $0.81$. The bottom center panel compares the full DM distributions. The reconstructed DM distribution also closely matches the true distribution, agreeing in both mean (reconstruction $0.86$ vs true $0.85$ in $\log_{10}\mathrm{DM}$) and width (reconstruction $0.31$ vs true $0.31$ dex). The model accurately reconstructs both individual sightlines and the overall statistics of the gas field. As expected, the model loses reconstruction fidelity at small scales. The input data (stellar mass-weighted galaxy density) does not contain significant information about the gas distribution at $k > 1\,h\,\rm Mpc^{-1}$, so the model must rely on its prior. This can be seen in the bottom right panel of Figure~\ref{fig:input_output}. The correlation between the reconstructed box and the true box falls at small scales, with nearly zero correlation near the spatial Nyquist frequency. For this reason, we consider the current algorithm as a method of reconstructing the cosmic web and not halo gas.
\figRank

A key benefit of probabilistic models like ours is that they produce a posterior distribution, not just a mean reconstruction. DM maps from the mean reconstruction tend to wash out small-scale structure, because in patches with low input galaxy density the model is uncertain and the mean averages over many possibilities. Individual samples from the reconstruction posterior are more visually similar to the true gas, because the learned prior demands the narrow filaments and low-density halos it has seen during training. 

To validate the model's probabilistic predictions, Figure~\ref{fig:patch_hist_box9} shows the posterior distribution of DM at three representative locations: a void, a filament, and a dense node. Because the model provides a distribution of voxel values rather than a point estimate, uncertainty quantification on FRB DMs follows naturally. Each posterior sample integrates to a single DM per voxel, so the spread across samples directly gives the predicted uncertainty on that sightline's DM. The true DM value consistently falls within the predicted distribution, and the posterior is, qualitatively, neither overconfident (a sharp peak that could miss ground truth) nor uninformative (spanning an unreasonably wide range).

\figCigar

Next, we assess the model's ability to preserve the relative ordering of DM values. We computed the rank correlation between true and reconstructed dispersion measures, finding a Spearman coefficient of $\rho = 0.857$. Figure \ref{fig:rank} shows a 2D histogram comparing the rank of true DM values to the rank of reconstructed DM values across all lines of sight in a representative test box. The concentration of density along the diagonal (white dashed line) indicates that the model accurately preserves the relative ordering of DM values from lowest to highest. This demonstrates that even where the absolute DM amplitude deviates from the truth, the model correctly identifies which regions are over/under-dense, confirming that it captures the relative structure of the cosmic web. This is the property that matters in practice. The absolute DM scale is set by the total amount of gas in the volume, $\Omega_b$, which is a known value, so an offset in the overall normalization can be calibrated out (Section~\ref{sec:discussion}).

\figFeedback

Figure~\ref{fig:pk} explores how the model responds to different feedback scenarios. We fed the same galaxy field into the model three times, with the same cosmology and the same initial noise, changing only the suppression spectrum, corresponding to high, correct, and low feedback. With high feedback, the model produces a smoother, more diffuse gas, as expected when strong feedback has blown gas out of dense structures into the voids. With the correct spectrum for the test box, the reconstruction recovers the true gas distribution. With low feedback, the model produces a more concentrated distribution with tighter filaments and denser nodes. These trends follow the physically expected direction as we vary $sP(k)$, which shows the model responds to the feedback conditioning.
Figure~\ref{fig:sigma} quantifies this response across the full test set. For each of the ten test boxes we compare the width $\sigma$ of the $\log_{10}\mathrm{DM}$ distribution, averaged over 100 posterior samples, between the high- and low-feedback reconstructions. In all ten boxes the low-feedback reconstruction yields the broader DM distribution, as expected when weaker suppression leaves the gas more concentrated in filaments and nodes. 
Averaged over the test set the width increases by $\sigma_\mathrm{low} - \sigma_\mathrm{high} = 0.0163 \pm 0.0020$ dex, or $5.6\%$ of $\sigma_\mathrm{high}$, with $p = 1.9\times10^{-5}$ for a paired comparison across the ten boxes. All ten differences share the same sign, for which a sign test gives $p = 0.002$. 
The direction of the response is therefore systematic across the test set rather than a feature of the single box.

\figSigma

\subsection{Sensitivity to the Galaxy Selection}
\label{sec:minms}

\figMinMS

\figMinMSrk

\begin{deluxetable}{ccccc}
\tabletypesize{\footnotesize}
\tablecaption{Reconstruction quality as a function of the minimum stellar mass threshold. \label{tab:minms}}
\tablehead{
\colhead{$\log_{10} M_{\star,\mathrm{min}}$} & \colhead{$N_{\mathrm{gal}}$} &
\colhead{$k_{1/2}$} & \colhead{$\rho$} &
\colhead{$\sigma_{\rm post}$} \\
\colhead{$[M_\odot]$} & \colhead{} &
\colhead{$[h\,\mathrm{Mpc}^{-1}]$} & \colhead{} & \colhead{[dex]}
}
\startdata
9.0  & 2558 & 1.44 & 0.833 & 0.133 \\
9.5  & 1420 & 1.44 & 0.789 & 0.144 \\
10.0 & 714  & 1.19 & 0.734 & 0.152 \\
10.5 & 314  & 0.94 & 0.632 & 0.161 \\
\enddata
\end{deluxetable}

$M_{\star,\mathrm{min}}$ is a conditioning parameter in our pipeline. It is drawn log uniformly between $10^9$ and $10^{9.5}\,M_\odot$ during training. This parameter is important for eventually matching the training data to the completeness of observed galaxy surveys. To quantify how much the fidelity of our reconstructions depends on it, we ran inference with different $M_{\star,\mathrm{min}}$ thresholds ranging from $10^9$ to $10^{10.5}\,M_\odot$ (Figure~\ref{fig:minms}). Thresholds above $10^{9.5}\,M_\odot$ lie outside the range sampled during training. We drew 10 posterior samples per held-out box and calculated the Spearman rank correlation ($\rho$) between the true and reconstructed DM, $\sigma_{\rm post}$ the spread of $\log_{10}\mathrm{DM}$ per sightline across posterior samples, and the cross-correlation coefficient $r(k)$ between the true and reconstructed gas fields. The average number of galaxies per box drops from 2558 to 314 galaxies, though the sparsest threshold is not uniformly sparse. At $10^{10.5}\,M_\odot$ the number of galaxies per box ranges from 16 to 915. The scale at which the correlation coefficient falls to 0.5 moves from $k = 1.44$ to $0.94\,h\,\mathrm{Mpc}^{-1}$ as galaxy number density drops in the box (Figure~\ref{fig:minms_rk}). This indicates the resolution coarsens and we lose small-scale detail. However, our small-scale fidelity is limited at all thresholds and by $k = 4\,h\,\mathrm{Mpc}^{-1}$, $r$ is below $0.15$ regardless of the threshold, so in that context the loss is minimal. The rank ordering degrades smoothly from $\rho = 0.833$ to $0.632$ as the number of galaxies per box decreases. The width of the posterior, averaged over all sightlines and boxes, increases by 21\% as the $M_{\star,\mathrm{min}}$ threshold is raised, suggesting the model registers when it is lacking information. The error between the reconstruction and the truth grows faster, by 39\% over the same range. The posterior is about 15\% too narrow at the lowest threshold and about 30\% too narrow at the highest. The model widens its uncertainty in the right direction as the catalog thins, but not by enough. Table~\ref{tab:minms} shows the metrics discussed in this section averaged over all 10 CAMELS50 held-out boxes for all thresholds. 

\subsection{Scaling to Larger Volumes}
Real FRBs sit at cosmological distances, far beyond the $50\,h^{-1}\,\mathrm{Mpc}$ of a single training box, so a useful reconstruction has to cover much longer sightlines. To test whether the model generalizes beyond the training distribution, we applied it to TNG300-1, a simulation box $\sim$4 times larger than the CAMELS50 training data (205 vs 50 $h^{-1}$ Mpc on a side). We gridded the TNG300-1 volume onto $524^3$ voxels so that the physical voxel size matched the $50\,h^{-1}\,\mathrm{Mpc}/128$ resolution of CAMELS50, and preprocessed it with the same pipeline as our training set. Our flow matching model is fully convolutional and therefore can be applied to a larger input volume with no change at all to the network or its trained weights. Figure \ref{fig:tng300} shows that the model successfully reconstructs the DM distribution across the larger box. The reconstructed DM histogram closely matches the true distribution, which demonstrates that the model generalizes in scale rather than memorizing structure at the size of the training box, and can scale to significantly larger volumes than those used for training. We note that TNG300-1 shares the IllustrisTNG galaxy formation model with our CAMELS50 training set, so this is a test of generalization in box size and resolution rather than in the underlying physics. The reconstruction underpredicts DM by approximately $0.17$ dex on average; in Figure~\ref{fig:tng300} the reconstructed and true DM are median matched as a demonstration. In practice, the absolute DM normalization can be anchored through the known value of $\Omega_b$ from cosmology, which allows us to calibrate out this offset and extract accurate DM values.

Because the model is fully convolutional, reconstructing a larger volume amounts to running it on a larger input, as the TNG300-1 test above shows; this is how the method reaches the volumes needed for cosmological sightlines. 
Separately, to illustrate what an extended reconstructed sightline looks like, we stitched together 10 CAMELS50 test boxes into a single line of sight extending to $\sim500\,h^{-1}\,\mathrm{Mpc}$. Figure~\ref{fig:cigar} compares the true and reconstructed gas along the stitched volume, with an example FRB sightline marked and the star showing the source.

\section{Discussion}
\label{sec:discussion}
We introduce a learning-based probabilistic 3D reconstruction of the ionized gas field from galaxy density fields. The reconstruction reproduces the true gas distribution, both along individual lines of sight and in its overall statistics, and it responds to the feedback conditioning through $sP(k)$. Because the model returns a full posterior rather than a single map, it also gives us the uncertainty on the gas, and on quantities we derive from it, including the dispersion measure. In what follows we discuss how this reconstruction opens a new path to constrain feedback from FRBs, how we anchor the absolute DM, how the method compares to previous work, what other areas our method may be useful in, and its current limitations.

\figTNG

Figure~\ref{fig:pk} shows that the reconstruction changes in a physical way when we change $sP(k)$. Stronger feedback gives smoother gas distributions, weaker feedback gives tighter filaments and denser nodes. We can turn this around and use it to constrain feedback. For a sample of FRBs with measured DMs and an observed foreground galaxy field, our model could be used to reconstruct over a range of input $sP(k)$ and, for each one, predict the distribution of DM. The $sP(k)$ whose predicted DM distribution is preferred by the data can be inferred, and feedback models that cannot produce that $sP(k)$ could be ruled out. The comparison would be made at the level of the FRB sample, not a single sightline. One FRB gives a single DM, which constrains only how much gas lies along that line of sight, whereas $sP(k)$ is a scale-dependent curve. 

For both TNG300-1 and on the CAMELS50 test boxes, the reconstructed DM distribution has a slightly different mean than the true one. On the ten CAMELS50 test boxes the reconstruction over-predicts the mean $\log_{10}\mathrm{DM}$ in every box, by $0.003$ to $0.058$ dex; on TNG300-1 it under-predicts it, by $0.17$ dex. This offset is in the overall density normalization, not in the structure. In Figure~\ref{fig:tng300}, the gas reconstructions have been normalized to have the same mean, erasing the offset in the DM histogram. The rank correlation between true and reconstructed DM is high ($\rho = 0.857$, Figure~\ref{fig:rank}), so the reconstruction still recovers which sightlines are over/under-dense relative to one another. The same is true of the dynamic range where the relation between reconstructed and true DM is slightly shallower than unity (Section~\ref{sec:results}), so the highest-DM sightlines are slightly under-predicted and the lowest somewhat over-predicted. Because both effects are monotonic, neither disturbs the ordering of sightlines, and both are absorbed by the same calibration. The absolute scale of the DM is set by the cosmic baryon density $\Omega_b$, which is known from cosmology. We anchor the normalization to $\Omega_b$, which removes the offset and yields accurate absolute DM. When our method is applied to real data in future work, we recommend normalizing the reconstructions to have the empirical $\Omega_b$ value.

Our efforts to reconstruct FRB sightlines in hopes of ruling out feedback models and constraining cosmological and astrophysical parameters are in line with existing surveys such as FLIMFLAM \citep{Khrykin_2024} which constrained the IGM and CGM baryon fractions along 8 FRB sightlines by taking the spectra of all the foreground galaxies. They used an explicit Bayesian inference with a forward model, whereas our likelihood is learned from full hydrodynamic simulations so it inherits the feedback physics. The machinery differs, but all the subsequent science of constraining the parameters and partitioning DM contributions is similar. However, our model makes the power spectrum explicit, rather than quantities like $f_{\rm IGM}$. 

A natural alternative to a learned reconstruction is the halo method: use a halo profile, paint halos over galaxies from observation, and integrate the mass of halos. This works well for dark matter but most of the gas is not inside halos and is between them. For FRBs, that diffuse gas outside the halo profile is the dominant component of the dispersion measure \citep{connor2024}. Our reconstruction is field level and makes no assumptions about halos. A hybrid approach could be used to retain the small scale gas structure that our method misses.

\citet{Kvasiuk2024} and \citet{Krywonos2026} map galaxies onto electron density with a graph and convolutional network. Both studies return one field per input. \citet{Kvasiuk2024} argue that a deterministic reconstruction is sufficient for many cosmological applications, and that producing a full posterior distribution is computationally prohibitive. While that is true for their target application of kSZ velocity reconstruction, the situation is different for FRBs. The dispersion measure of an FRB is a single measurement. Knowing how likely that measurement is will enable us to (i) draw conclusions about the host DM, (ii) identify the outlier FRBs with unusually high or low DMs so that they do not bias our feedback constraints or other analyses, and (iii) treat the FRBs sightlines with tighter posteriors as more reliable than those passing through sparsely sampled regions and wide posterior distributions. Together, these mean that a given constraint on feedback can be reached with a smaller sample of FRBs than a point estimate would require. 

In addition to feedback and cosmological inference, 3D gas reconstructions are useful for other FRB science cases. For example, the host contribution to 
total DM is an important question for understanding FRB progenitor environments. Large contributions from the ISM and local nebulae are expected if FRBs are borne out of young stellar populations; low host DM is expected for FRBs with high galactic offsets, 
older stellar populations. To infer host DM, we typically subtract off $\left < \rm DM_{cos}(z) \right >$, agnostic about the large-scale 
structure in FRB foregrounds. Our method provides more accurate per-sightline $DM_{cos}(z)$ conditioned on the observed galaxy distribution. 

Beyond FRBs, our model is useful for observables 
that are concerned with the baryon large-scale structure. The flow matching 
approach developed here was originally built for cosmological velocity reconstruction.
The kinematic Sunyaev–Zeldovich (kSZ) effect is a velocity-weighted integral of exactly the $n_e$ field the model outputs. The model could be reconfigured to output both gas density and velocity, consistently. A per-object 3D posterior on $n_e$ and velocity  
offers kSZ templates for comparing against CMB and galaxy surveys. Joint inference could be done on kSZ, galaxies, and FRB DM observations, using our method. The model could output both temperature and gas density to enable tSZ estimates in a consistent way. One limitation of our model for the SZ effects is the fidelity of reconstructed small scales, so effort would need to go into accurately recovering structure at the scale of halos. Another application is a baryonic feedback correction for weak lensing. Cosmic shear is limited by knowledge of the baryon distribution, and our reconstructions could be used to forward-model weak lensing observations conditioned on feedback and cosmology. In this case, it may be useful to jointly output both gas density and dark matter density. Again, such maps could be used for inference on weak lensing shear maps and FRBs simultaneously, when large future FRB samples are available. Finally, diffuse emission from filaments and cluster outskirts is difficult to model, but stacking analyses can be used \citep{Tanimura_2020}. FRBs and X-rays both have the ability to measure this gas and the reconstructions presented here make probabilistic predictions for those observables. 

We have not yet tested the method on real galaxy survey data, opting to leave that to a future effort in which observational selection effects can be carefully considered.
The algorithm has been verified on simulated data under a range of conditions. Our current training and validation data assume uniform 
galaxy sampling and redshift-independent thresholds on stellar mass. While our galaxy number densities are comparable to 
existing spectroscopic galaxy surveys, we expect model 
mismatch between training and validation data to degrade 
reconstruction quality unless training data is forward modeled 
to match specific surveys. 

Finally, we wish to note limitations of the current method and future approaches to resolving them. The minimum stellar mass threshold that defines the galaxy sample is a free choice in our pipeline and is not tied to a specific survey. We pass it to the model as a conditioning parameter and marginalize over it at inference, but the reconstruction still depends on this choice, and we have not calibrated it against a real survey selection function. Relatedly, the model inherits the assumptions and systematics of the CAMELS IllustrisTNG simulations, which can bias inference if those assumptions are inaccurate. Although CAMELS parameters span a broad range by design, the inferred parameters are still limited to the simulation's prior range, which is a caveat of all simulation based inference models. The model is trained and evaluated at $z = 0$, while real FRB sightlines extend across a range of redshifts; applying the method at cosmological distances requires accounting for redshift evolution of gas. Finally, the input data (galaxy positions) does not contain information about the $k \gtrsim 1\,h\,\rm Mpc^{-1}$ gas distribution, forcing the model to rely on its prior. For applications that require halo gas, a hybrid halo-model/SBI approach must be taken or training data must include higher resolution gas fields. We consider each of these solvable in principle, but our current work is concerned with establishing the method as a viable path towards probabilistic baryon reconstruction. Calibrating survey selection functions will be required when our model is applied to real data. Threats of overfitting to our training data can 
be reduced by including new simulations (e.g. CAMELS Astrid, FLAMINGO) and a wider redshift range. 

\section{Conclusions}
\label{sec:conclusion}
We have introduced a flow matching model for reconstructing the 3D distribution of cosmic baryons from galaxy survey data, conditioned on cosmological parameters and the baryon suppression spectrum. The model learns to predict the posterior distribution of gas density fields given stellar mass-weighted galaxy density. 
The model is trained on the CAMELS50 (IllustrisTNG) simulation suite. We validated the model on test boxes, demonstrating accurate reconstruction of both the spatial structure and statistical distribution of the gas field, as measured by dispersion measure along lines of sight. The model generalizes to larger box sizes, successfully reconstructing DM distributions in TNG300-1. By conditioning on the baryon suppression spectrum, the model captures the integrated effects of feedback processes, allowing for future applications in constraining feedback models from FRB observations. This framework opens a new avenue for mapping the cosmic baryons and leveraging FRBs as probes of large-scale structure and galaxy evolution.

Future work will explore the sensitivity of reconstructions to different galaxy selection functions, training on multiple simulation suites to mitigate model dependence, and incorporating cosmological parameter uncertainties into the conditioning framework. Despite these caveats, our results demonstrate that flow matching models can effectively learn complex 3D distributions of cosmic baryons from galaxy observations, providing a powerful new tool for interpreting FRB measurements and probing the cosmic web.

\begin{acknowledgments}
We thank Shy Genel for help with the CAMELS simulations and useful discussions. M.H. and L.C. acknowledge support from the U.S. National Science Foundation Astronomy and Astrophysics Research Grants program under award AST-2508734. 
A.A. thanks the LSST-DA Data Science Fellowship Program, which is funded by LSST-DA, the Brinson Foundation, the WoodNext Foundation, and the Research Corporation for Science Advancement Foundation; her participation in the program has benefited this work.
The computations in this paper were run on the FASRC Cannon cluster supported by the FAS Division of Science Research Computing Group at Harvard University. 
\end{acknowledgments}
\clearpage

\bibliography{ref}
\end{document}

%% file: figures.tex
\newcommand{\figOverview}{%
  \begin{figure*}
    \centering
    \hspace*{-0.5cm}\includegraphics[width=1.05\linewidth]{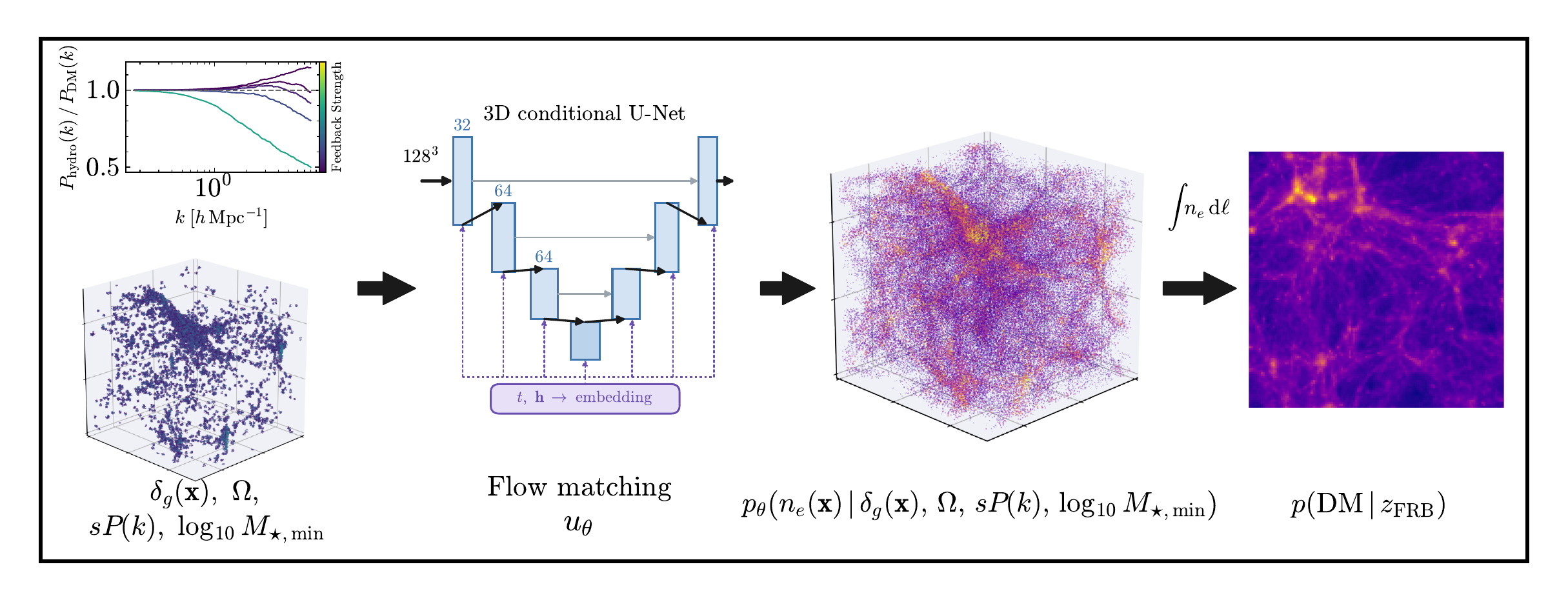}
    \caption{Overview of our method for reconstructing the 3D distribution of cosmic baryons. The flow matching model learns to reconstruct the gas density field conditioned on galaxy density field $\delta_g(\mathbf{x})$, minimum stellar mass threshold $\log_{10}M_{\star,\mathrm{min}}$ of the galaxy sample, the cosmological parameters $\mathbf{\Omega}$, and
    power spectrum suppression $sP(k)$. The output is a gas density field that can be integrated to predict FRB dispersion measures. The model is probabilistic, allowing sampling from a reconstruction posterior.
    }
    \label{fig:overview-fig}
  \end{figure*}
}

\newcommand{\figInput}{
  \begin{figure*}
    \centering
    \includegraphics[width=1.\linewidth]{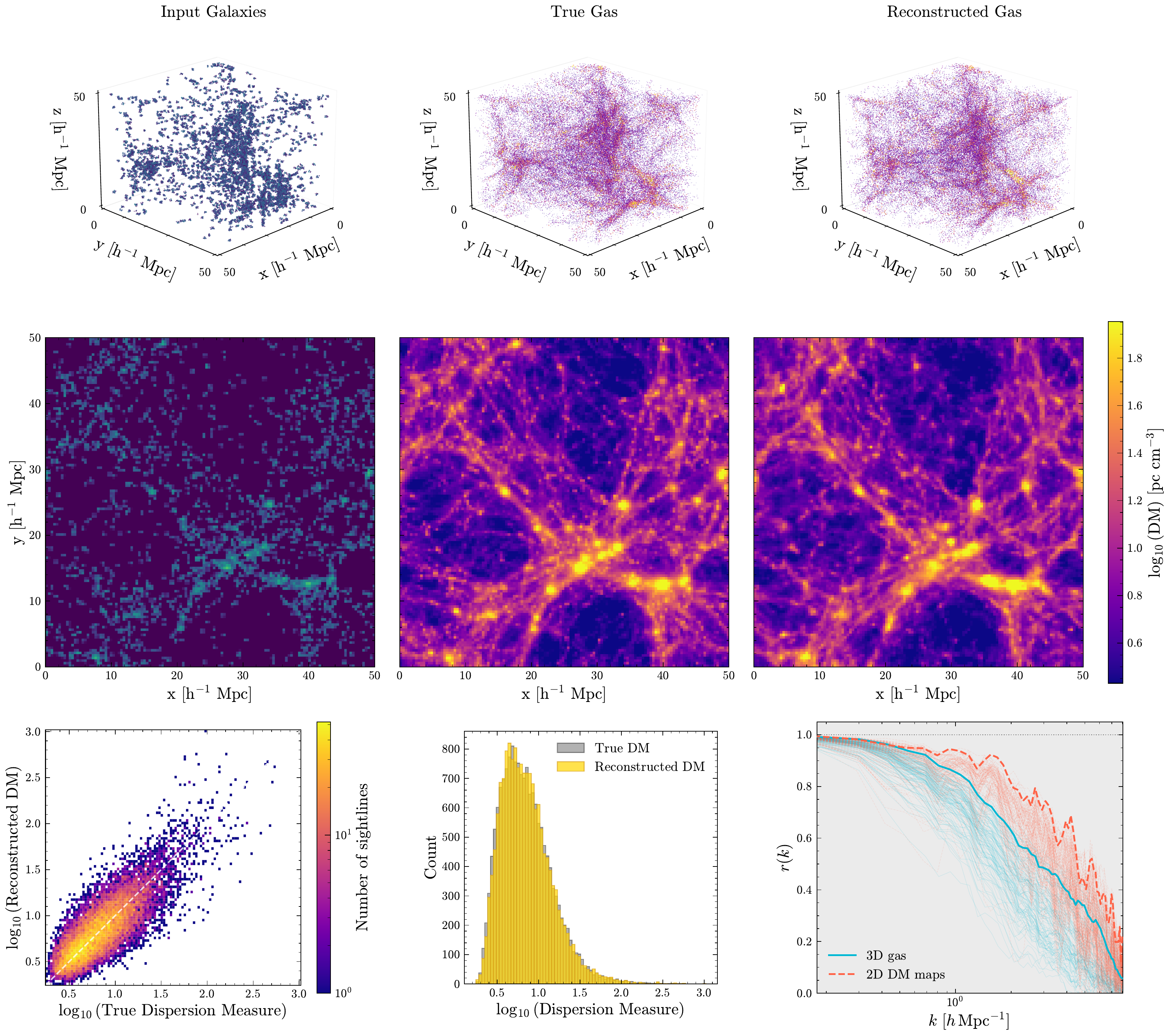}
    \caption{\textbf{Top row:} From left to right, the 3D input galaxy density field, 3D true field, and 3D reconstructed output gas field. \textbf{Middle row:} 2D map of the input galaxy density field projected along the line of sight (left), true dispersion measure (center) and reconstructed dispersion measure (right) computed by summing gas density along the line of sight. \textbf{Bottom row:} On the left, 2D histogram of the model's DM for a single posterior sample against the true DM across all lines of sight; the dashed line marks perfect 1:1 agreement. In the center, histogram comparing the distribution of true DM values (gray) to reconstructed DM values (gold). On the right, correlation coefficients for the 3D gas reconstruction and the inferred DM maps are shown in cyan and red, respectively, with darker curves corresponding to the mean reconstruction and lighter curves corresponding to individual posterior samples.}
    \label{fig:input_output}
  \end{figure*}
}

\newcommand{\figPatchHist}{%
  \begin{figure}
    \centering
    \includegraphics[width=\linewidth]{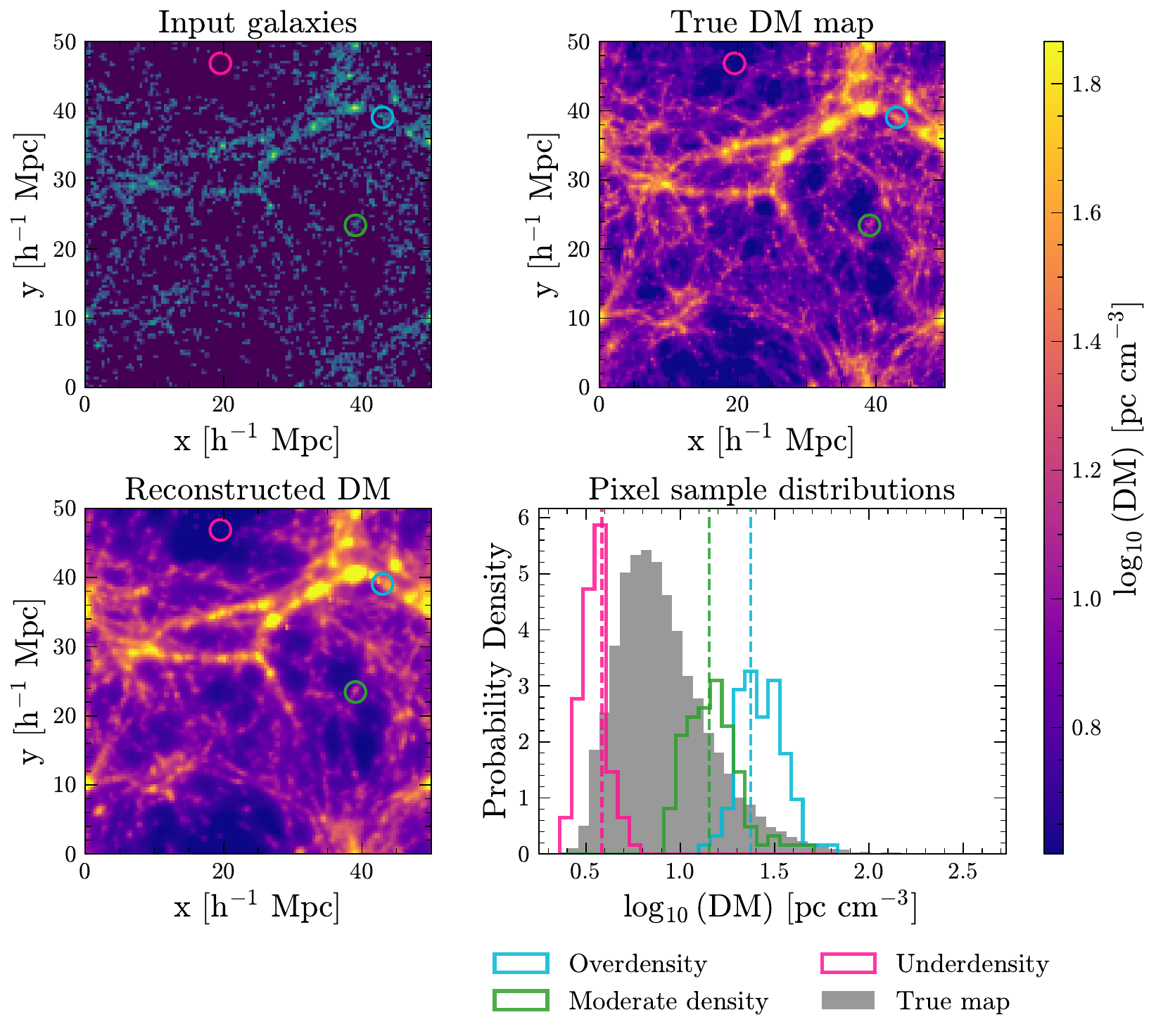}
    \caption{\textbf{Top left:} Input galaxy density field. \textbf{Top right:} True simulation DM map. \textbf{Bottom left:} Dispersion measure map of the model's posterior mean reconstruction. \textbf{Bottom right:} Probability density of DM values at three representative sightlines marked in the maps, spanning different cosmic environments (overdensity, moderate density, and underdensity). At each location, the colored distribution shows 100 independent posterior samples drawn from the flow matching model; the dashed vertical line marks the true DM value at that pixel; the gray distribution labeled ``True map'' shows the overall DM distribution across the full field for reference. The model's posterior samples cluster around the true DM value across all three environments, demonstrating the model's uncertainty estimation abilities.}
    \label{fig:patch_hist_box9}
  \end{figure}
}

\newcommand{\figFeedback}{%
  \begin{figure*}
    \centering
    \includegraphics[width=0.95\linewidth]{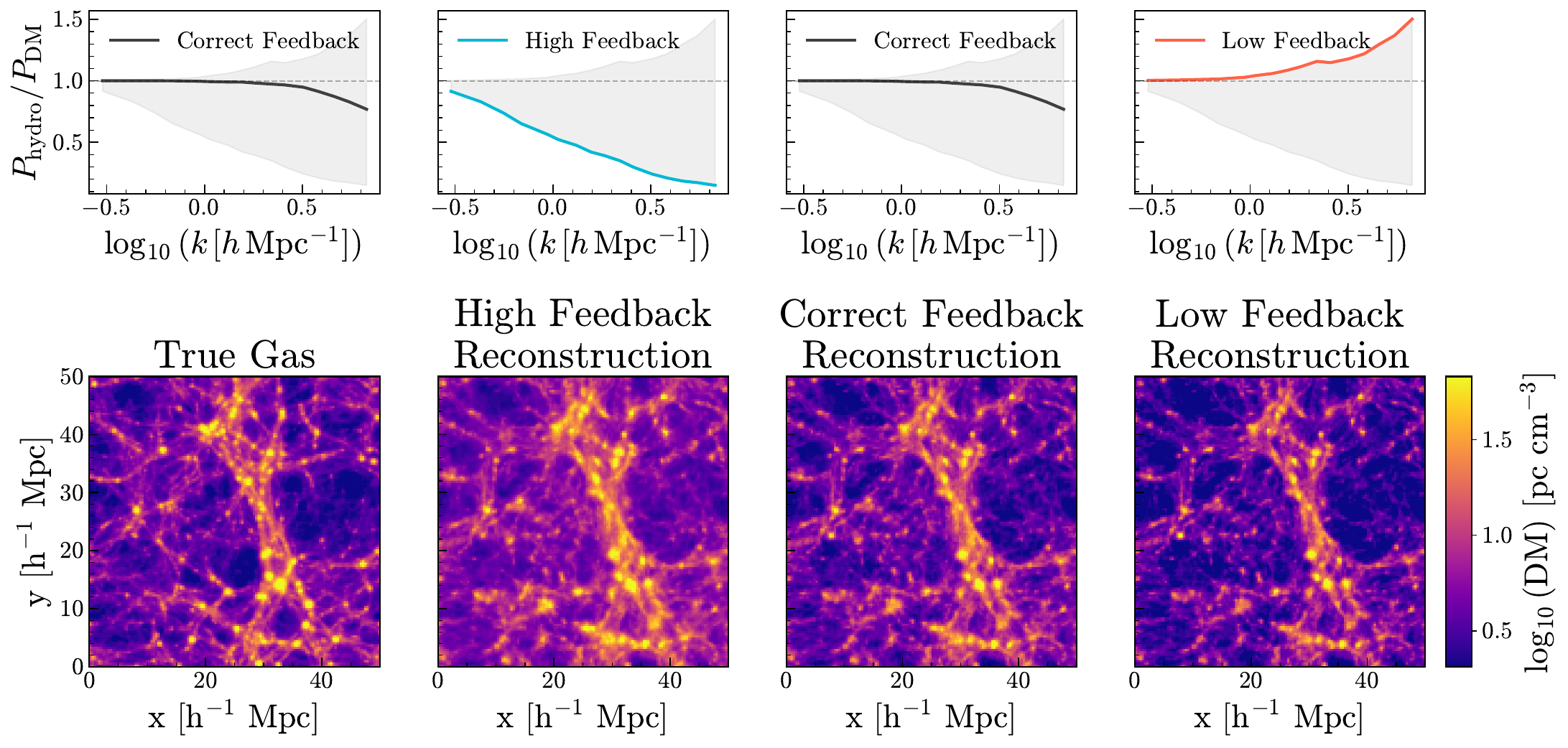}
    \caption{The response of the reconstruction to the feedback conditioning, for one test box. \textbf{Top row:} the baryon suppression spectrum $sP(k)$ the model was conditioned on in each column. The light grey shows the $sP(k)$ range of the simulation. The box's own (correct) $sP(k)$ is in black. The strongest suppression (high feedback) is in cyan, and the weakest (low feedback) is in red. \textbf{Bottom row, left to right:} the true DM map of the box, followed by the sample DM maps reconstructed under high, correct, and low feedback. All four columns share the same galaxy input, cosmology, and initial noise, so the only thing that changes across the reconstructions is $sP(k)$. The baryon suppression spectrum encodes both cosmological parameters (determining total baryon abundance) and feedback strength (determining baryon redistribution). Strong feedback gives a more diffuse gas distribution, while weak feedback gives gas concentrated in filaments and nodes. The correct feedback panel matches the true field, validating that the model successfully learns the relationship between the suppression spectrum and gas structure.}
    \label{fig:pk}
  \end{figure*}
}
\newcommand{\figTNG}{%
  \begin{figure*}
    \centering
    \includegraphics[width=\linewidth]{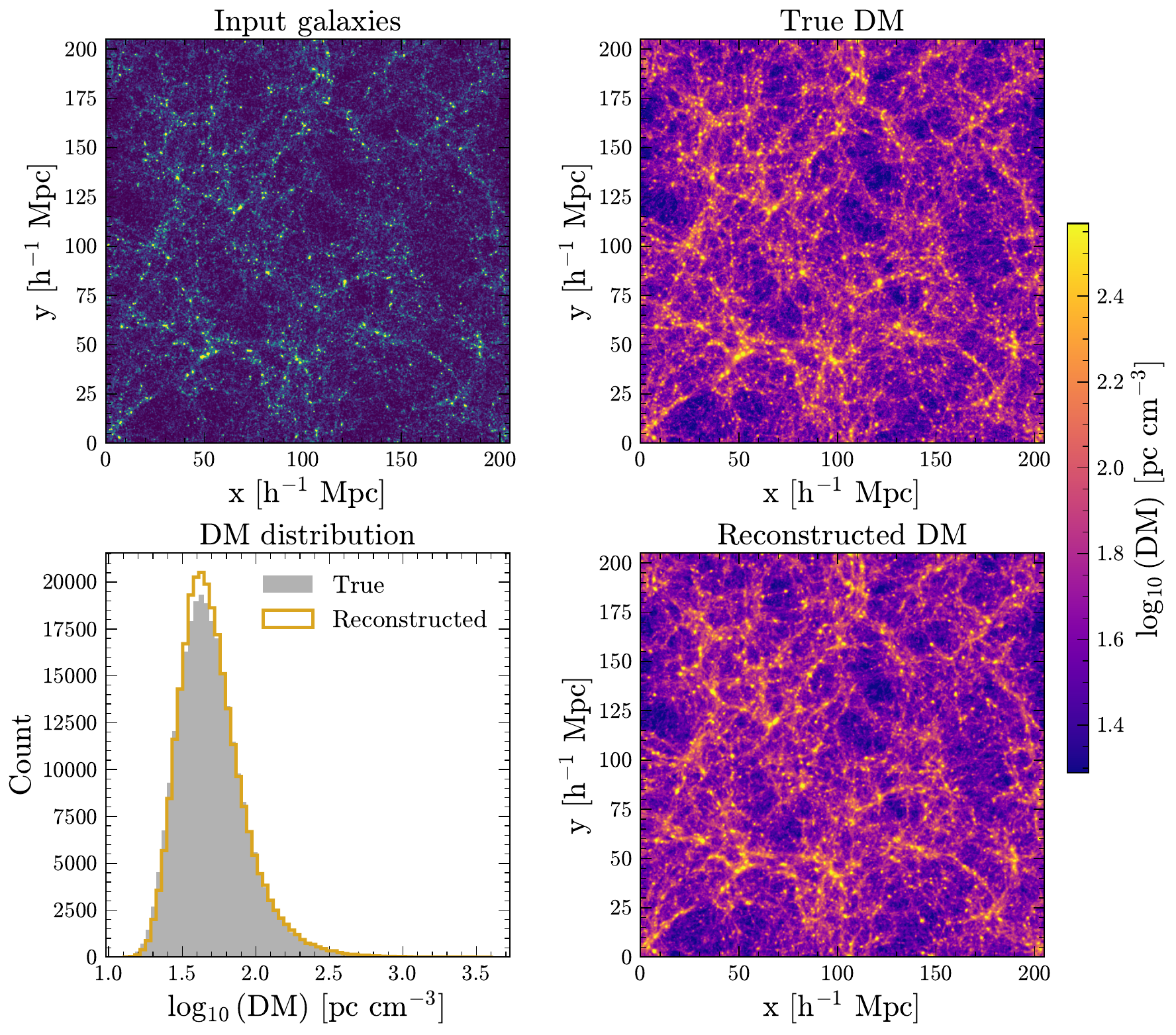}
    \caption{\textbf{(Top left)} Input galaxy density field; \textbf{(top right)} true DM map from a TNG300-1 simulation box of size $(205\,h^{-1}\,\mathrm{Mpc})^3$; \textbf{(bottom right)} the model's reconstructed DM. The model was trained exclusively on CAMELS50 boxes of $(50\,h^{-1}\,\mathrm{Mpc})^3$, making this a $4\times$ scaling test in each side. \textbf{(Bottom left)} Histogram comparing the distribution of true DM values (gray) to reconstructed DM values (gold) across the entire TNG300-1 box. The reconstructed DM median has been calibrated to match the median of the true DM, correcting an overall $\sim0.17$ dex underprediction by the model.
}
    \label{fig:tng300}
  \end{figure*}
}

\newcommand{\figCigar}{%
  \begin{figure*}
    \centering
    \includegraphics[width=0.98\linewidth]{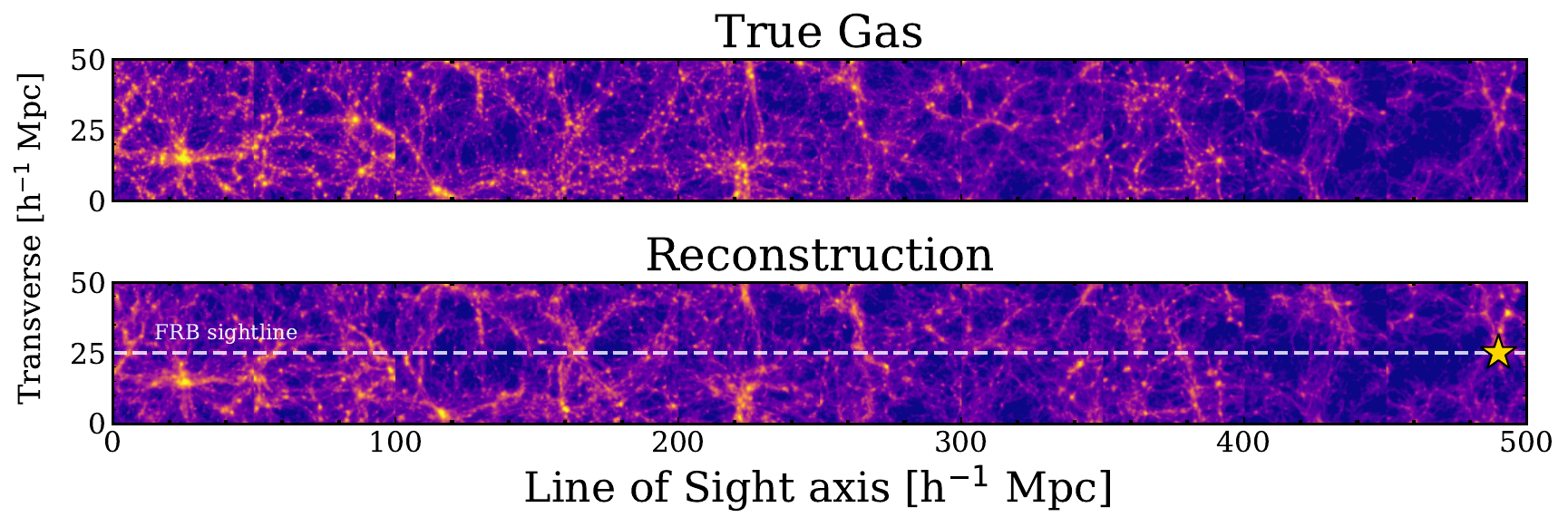}
    \caption{An illustration of what an extended reconstructed sightline looks like. 10 CAMELS50 test boxes are tiled along the line-of-sight axis to construct an extended sightline spanning $\sim500\,h^{-1}\,\mathrm{Mpc}$. \textbf{Top:} True gas density field from the simulation. \textbf{Bottom:} Model reconstruction of the gas density field. The white dashed line marks an example FRB sightline and the star marks the FRB source.}
    \label{fig:cigar}
  \end{figure*}
}

\newcommand{\figRank}{%
  \begin{figure}
    \centering
    \includegraphics[width=\linewidth]{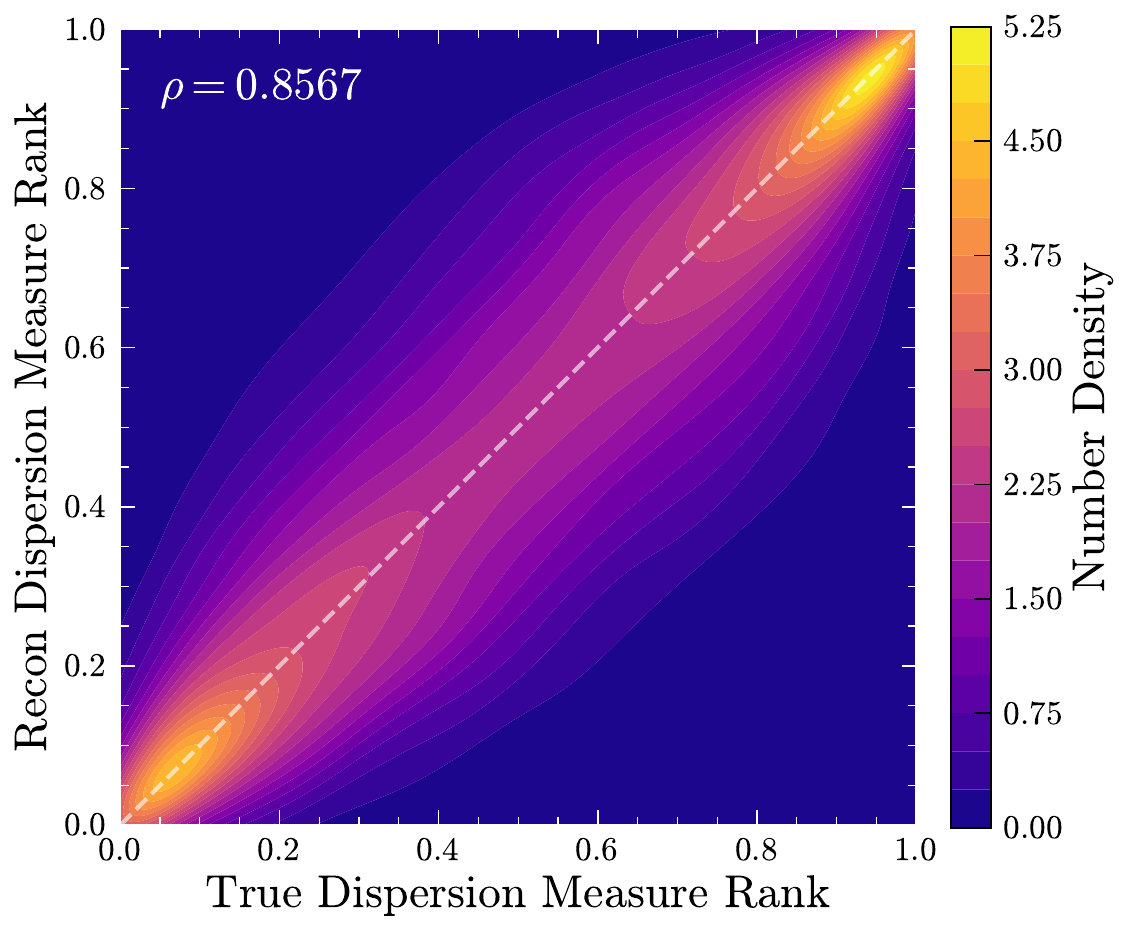}
     \caption{
    2D histogram comparing the rank of true DM values (x-axis) to the rank of reconstructed DM values from the model's posterior mean (y-axis), with both axes normalized to $[0, 1]$. The white dashed line indicates perfect 1:1 agreement. The color scale (purple to yellow) shows the number density of lines of sight at each rank pair. The concentration of points along the diagonal demonstrates that the model preserves the relative ordering of DM values across the full dynamic range, even where absolute predictions may deviate. The Spearman rank correlation is $\rho = 0.857$.
}
    \label{fig:rank}
  \end{figure}
}

\newcommand{\figSigma}{%
  \begin{figure}
    \centering
    \includegraphics[width=\linewidth]{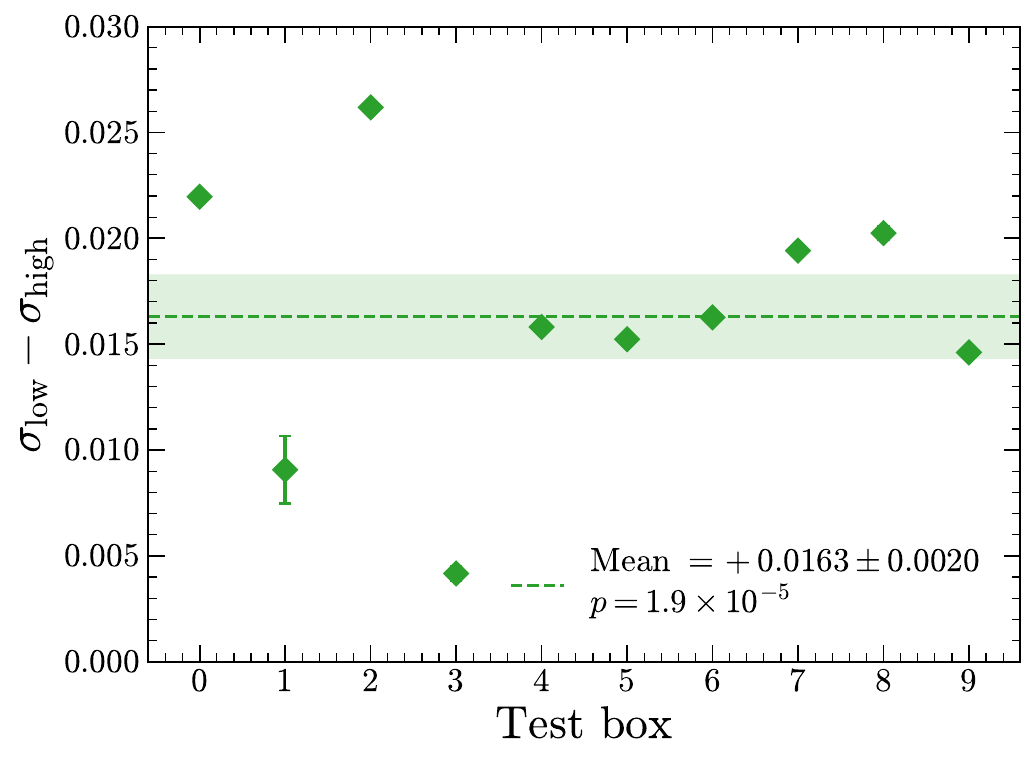}
    \caption{
    Green diamonds show the difference in the width $\sigma$ of the $\log_{10}\mathrm{DM}$ distribution, averaged over 100 posterior samples for high feedback and low feedback reconstructions, for each test box. The error bar on each diamond is the standard error over those 100 paired draws. The dashed line is the mean over the ten boxes, the shaded band is $\pm1$ standard error on that mean, computed from the scatter of the ten per-box values.  Reconstructions conditioned on weak baryonic suppression (low feedback) produce a broader DM distribution than those conditioned on strong suppression (high feedback), as expected. The difference is positive in all ten boxes, with a mean of $0.0163 \pm 0.0020$ dex ($5.6\%$ of $\sigma_\mathrm{high}$), $p = 1.9\times10^{-5}$ for a paired comparison across boxes.
}
    \label{fig:sigma}
  \end{figure}
}

\newcommand{\figMinMS}{%
  \begin{figure*}[!t]
    \centering
    \includegraphics[width=\linewidth]{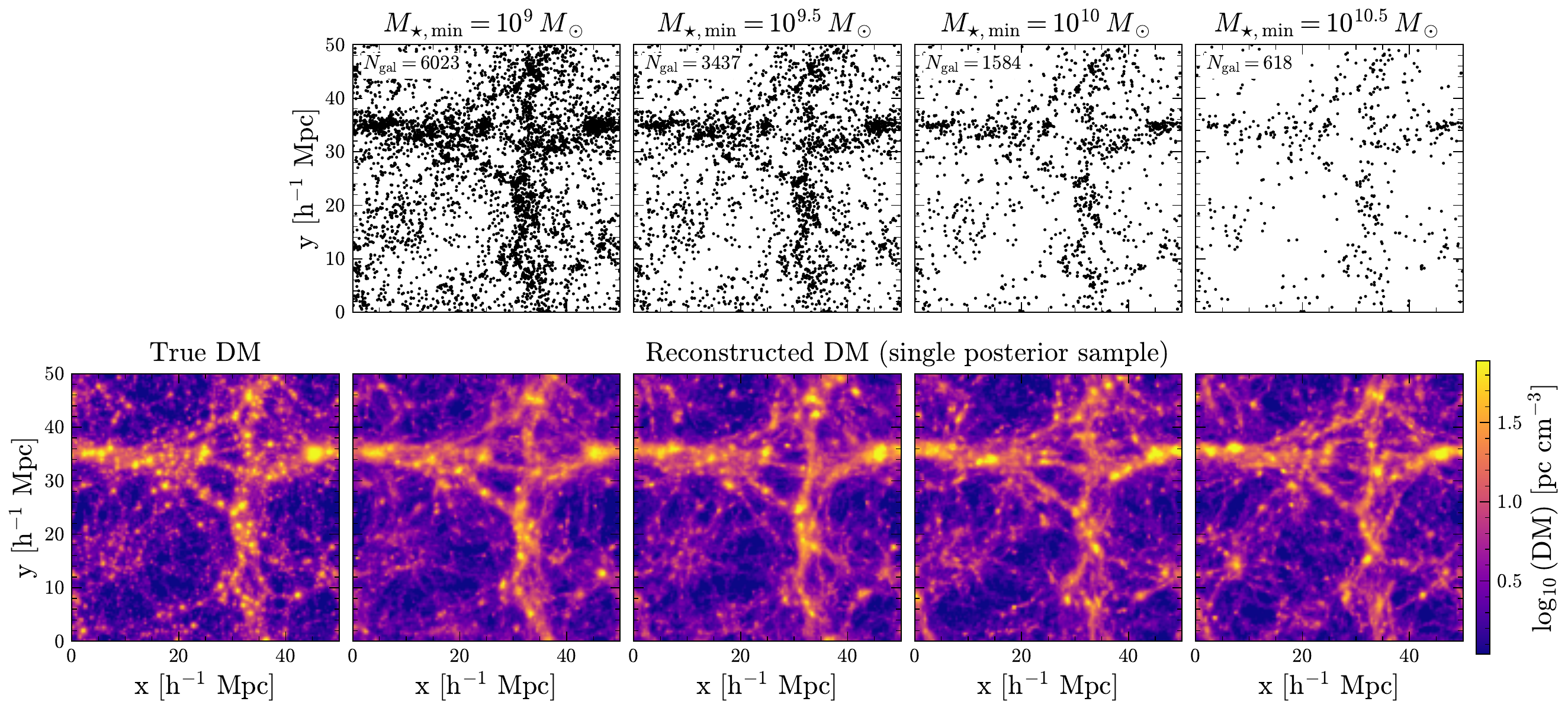}
    \caption{
    Sensitivity of the reconstruction to the galaxy selection, for a single CAMELS50 test box (box 0). \textbf{Top row:} the galaxy field at four values of the minimum stellar mass threshold, $M_{\star,\mathrm{min}} = 10^{9}$, $10^{9.5}$, $10^{10}$ and $10^{10.5}\,M_\odot$, with the number of galaxies $N_{\mathrm{gal}}$ in each panel. \textbf{Bottom row:} the true DM map for the same box (left), followed by a single posterior sample of the reconstructed DM at each threshold, in the same column order as the top row. The two highest thresholds lie above the $10^{9}$ to $10^{9.5}\,M_\odot$ range sampled during training.
}
    \label{fig:minms}
  \end{figure*}
}

\newcommand{\figMinMSrk}{%
  \begin{figure}[!htb]
    \centering
    \includegraphics[width=\linewidth]{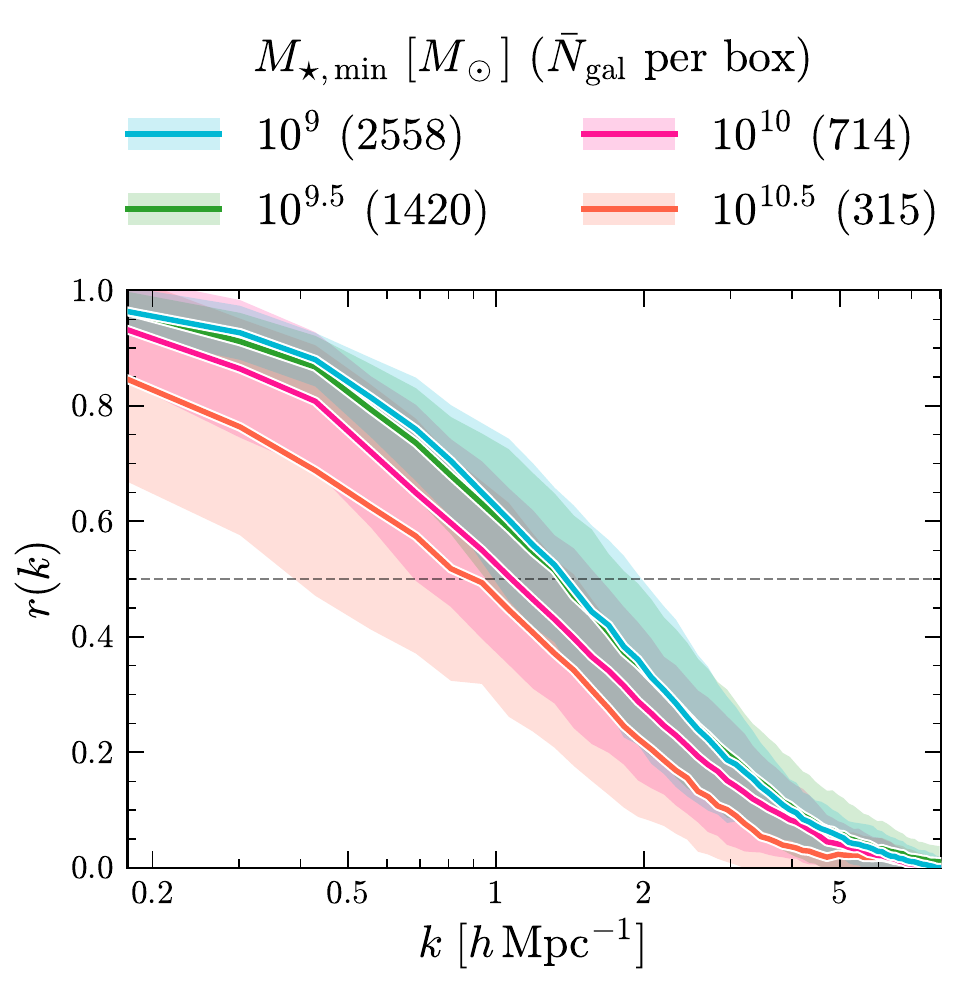}
    \caption{
    Cross-correlation coefficient $r(k)$ between the true and reconstructed 3D gas fields as a function of wavenumber, for each minimum stellar mass threshold. Each curve is the mean over the ten CAMELS50 test boxes with ten posterior samples per box, and the shaded band is the box to box scatter. The legend gives the mean number of galaxies per box at each threshold. The dashed line marks $r = 0.5$, which defines the half-coherence scale $k_{1/2}$ listed in Table~\ref{tab:minms}.
}
    \label{fig:minms_rk}
  \end{figure}
}